\documentclass[twocolumn]{aastex63}
\usepackage{tabularx}
\usepackage{url}

\shorttitle{ALCS: Properties of X-ray- and Millimeter-selected AGNs}
\shortauthors{Uematsu et al.}

\graphicspath{{./}{figures/}}

\begin{document}

\title{ALMA Lensing Cluster Survey: Properties of
Millimeter Galaxies Hosting X-ray Detected Active Galactic Nuclei}

\correspondingauthor{Ryosuke Uematsu}
\email{uematsu@kusastro.kyoto-u.ac.jp}

\author[0000-0001-6653-779X]{Ryosuke Uematsu}
\affiliation{Department of Astronomy, Kyoto University, Kyoto 606-8502, Japan}

\author[0000-0001-7821-6715]{Yoshihiro Ueda}
\affiliation{Department of Astronomy, Kyoto University, Kyoto 606-8502, Japan}

\author[0000-0002-4052-2394]{Kotaro Kohno}
\affiliation{Institute of Astronomy, Graduate School of Science, The University of Tokyo, 2-21-1 Osawa, Mitaka, Tokyo 181-0015, Japan}
\affiliation{Research Center for the Early Universe, School of Science, The University of Tokyo, 7-3-1 Hongo, Bunkyo-ku, Tokyo 113-0033, Japan}

\author[0000-0002-9754-3081]{Satoshi Yamada}
\affiliation{Institute of Physical and Chemical Research (RIKEN), 2-1 Hirosawa, Wako, Saitama 351-0198}

\author[0000-0002-3531-7863]{Yoshiki Toba}
\affiliation{National Astronomical Observatory of Japan, 2-21-1 Osawa, Mitaka, Tokyo 181-8588, Japan}
\affiliation{Department of Astronomy, Kyoto University, Kyoto 606-8502, Japan}
\affiliation{Academia Sinica Institute of Astronomy and Astrophysics, 11F of Astronomy-Mathematics Building, AS/NTU, No.1, Section 4, Roosevelt Road, Taipei 10617, Taiwan}
\affiliation{Research Center for Space and Cosmic Evolution, Ehime University, 2-5 Bunkyo-cho, Matsuyama, Ehime 790-8577, Japan}

\author[0000-0001-7201-5066]{Seiji Fujimoto}
\affiliation{Cosmic Dawn Center (DAWN), Jagtvej 128, DK2200 Copenhagen N, Denmark}
\affiliation{Niels Bohr Institute, University of Copenhagen, Lyngbyvej 2, DK2100 Copenhagen Ø, Denmark}

\author[0000-0001-6469-8725]{Bunyo Hatsukade}
\affiliation{Institute of Astronomy, Graduate School of Science, The University of Tokyo, 2-21-1 Osawa, Mitaka, Tokyo 181-0015, Japan}

\author[0000-0003-1937-0573]{Hideki Umehata}
\affiliation{Institute for Advanced Research, Nagoya University, Furocho, Chikusa, Nagoya 464-8602, Japan}
\affiliation{Department of Physics, Nagoya University, Furo-cho, Chikusa-ku, Nagoya 464-8601, Japan}

\author[0000-0002-8726-7685]{Daniel Espada}
\affiliation{Departamento de F\'{i}sica Te\'{o}rica y del Cosmos, Campus de Fuentenueva, Edificio Mecenas, Universidad de Granada, E-18071, Granada, Spain}
\affiliation{Instituto Carlos I de F\'{i}sica Te\'{o}rica y Computacional, Facultad de Ciencias, E-18071, Granada, Spain}

\author[0000-0002-4622-6617]{Fengwu Sun}
\affiliation{Steward Observatory, University of Arizona, 933 N. Cherry Avenue, Tucson, 85721, USA}

\author[0000-0002-4872-2294]{Georgios E. Magdis}
\affiliation{Cosmic Dawn Center (DAWN), Jagtvej 128, DK2200 Copenhagen N, Denmark}
\affiliation{DTU-Space, Technical University of Denmark, Elektrovej 327, 2800, Kgs. Lyngby, Denmark}
\affiliation{Niels Bohr Institute, University of Copenhagen, Lyngbyvej 2, DK2100 Copenhagen Ø, Denmark}

\author[0000-0002-5588-9156]{Vasily Kokorev}
\affiliation{Kapteyn Astronomical Institute, University of Groningen, P.O. Box 800, 9700AV Groningen, The Netherlands}
\affiliation{Cosmic Dawn Center (DAWN), Jagtvej 128, DK2200 Copenhagen N, Denmark}
\affiliation{Niels Bohr Institute, University of Copenhagen, Lyngbyvej 2, DK2100 Copenhagen Ø, Denmark}

\author[0000-0003-3139-2724]{Yiping Ao}
\affiliation{Purple Mountain Observatory and Key Laboratory for Radio Astronomy, Chinese Academy of Sciences, Nanjing, China}

\begin{abstract}

We report the multi-wavelength properties of millimeter galaxies hosting X-ray detected active galactic nuclei (AGNs) from the ALMA Lensing Cluster Survey (ALCS). ALCS is an extensive survey of well-studied lensing clusters with ALMA, covering an area of 133 arcmin$^2$ over 33 clusters with a 1.2 mm flux-density limit of ${\sim}$60 $\mathrm{\mu Jy}$ ($1\sigma$). Utilizing the archival data of \textit{Chandra}, we identify three AGNs at $z=$1.06, 2.09, and 2.84 among the 180 millimeter sources securely detected in the ALCS (of which 155 are inside the coverage of \textit{Chandra}). The X-ray spectral analysis shows that two AGNs are not significantly absorbed ($\log N_{\mathrm{H}}/\mathrm{cm}^{-2} < 23$), while the other shows signs of moderate absorption ($\log N_{\mathrm{H}}/\mathrm{cm}^{-2}\sim 23.5$). We also perform spectral energy distribution (SED) modelling of X-ray to millimeter photometry. We find that our X-ray AGN sample shows both high mass accretion rates (intrinsic 0.5--8 keV X-ray luminosities of ${\sim}10^{\text{44--45}}\,\mathrm{erg\ s^{-1}}$) and star-formation rates (${\gtrsim}100\,M_{\odot}\,\mathrm{yr}^{-1}$). This demonstrates that a wide-area survey with ALMA and \textit{Chandra} can selectively detect intense growth of both galaxies and supermassive black holes (SMBHs) in the high-redshift universe.

\end{abstract}


\section{Introduction} \label{sec:intro}

The evolution of galaxies and the supermassive black holes (SMBHs) at their centers is one of the most important issues in modern astronomy. Many studies have shown the tight bulge-mass-to-SMBH-mass correlation in the local universe (e.g., \citealt{1998AJ....115.2285M,2003ApJ...589L..21M}), suggesting the co-evolution of SMBHs and the host galaxies. Nevertheless, the large dispersion in the bulge-mass-to-SMBH-mass ratio in high-redshift galaxies (e.g., see \citealt{2013ARA&A..51..511K} for a review) suggests complexity of the co-evolution scenario, and the evolution history of individual system is still unclear.

The averaged growth rate of galaxies and SMBHs reached a peak at $z$=1--3, which is often referred to as ``cosmic noon'' \citep{2014ARA&A..52..415M, 2014ApJ...786..104U}. Thus, galaxies hosting active galactic nuclei (AGNs) at these epochs are a key population to reveal the mechanisms of galaxy-SMBH co-evolution. Submillimeter and X-ray observations are powerful tools to study this population. This is because infrared radiation from star-formation activity at these redshifts is observed in millimeter band (submillimeter galaxies; SMGs), while X-ray observations can detect the obscured AGNs. The Atacama Large Millimeter/submillimeter Array (ALMA) has unprecedentedly high angular resolution and sensitivity in sub/millimeter wavelengths, and has been used extensively to study high-redshift universe. For example, \citet{2013ApJ...778..179W} found 10 X-ray counterparts to 99 SMGs of ALMA LABOCA E-CDFS Submillimeter Survey (ALESS; \citealt{2013ApJ...768...91H}), from which 8 sources were identified as AGNs; hereafter we refer to these X-ray AGNs as ``ALESS-XAGN'' sample. Moreover, \citet{2016ApJ...833...12R} found 6 X-ray AGNs from 16 SMGs of GOODS-S/ultra deep field (UDF) survey (\citealt{2017MNRAS.466..861D}); hereafter ``UDF-XAGN''. Furthermore, \citet{2018ApJ...853...24U} identified 13 X-ray AGNs in a sample consisting of 25 SMGs from the ALMA twenty-Six Arcmin$^2$ survey of GOODS-S At One-millimeter (ASAGAO; \citealt{2018PASJ...70..105H}), supplemented by the UDF survey (6 X-ray AGNs are the same as UDF-AGN); hereafter, we refer to those X-ray AGNs that are not in the UDF region as ``ASAGAO-XAGN''. In another case, \citet{2015ApJ...815L...8U} found 4 X-ray AGNs from 8 SMGs of SSA22 protocluster. Finally, \citet{2019MNRAS.487.4648S} identified 23 X-ray AGNs from 274 SMGs of ALMA SCUBA-2 UDS survey (AS2UDS; \citealt{2018ApJ...860..161S}) falling within the Chandra footprint; hereafter ``AS2UDS-XAGN". These sub/millimeter surveys only cover a limited survey parameter space (depth and area), however. In shallow surveys, the results are biased for intensively star-forming galaxies, whereas rare populations are missed in narrow area surveys. It is important to conduct deep {\it and} wide area surveys to carry out a complete census of the whole sub/millimeter populations.

Lensed fields are excellent targets to perform deep surveys efficiently. Using ALMA in cycle-6, our team has conducted an extensive survey in high magnification regions of 33 lensing clusters, called the ALMA lensing cluster survey (ALCS; Fujimoto et al., in prep.; Kohno et al., in prep.). The sample comes from the best-studied clusters of galaxies observed with the \textit{Hubble Space Telescope} (\textit{HST}) treasury programs, i.e., CLASH \citep{2012ApJS..199...25P}, HFF \citep{2017ApJ...837...97L,2020ApJS..247...64S}, and RELICS \citep{2019ApJ...884...85C}. The total survey area is 133 $\mathrm{arcmin}^2$ and the depth is $\sim$60 $\mu$Jy (1.2 mm, 1 $\sigma$). The ALCS achieves, after correcting for lensing, one of the widest and deepest millimeter surveys among ALMA unbiased surveys conducted to date (Fujimoto et al. in prep.).

In this paper, we investigate the properties of the three millimeter galaxies hosting X-ray detected AGNs in the ALCS, using the archival data of \textit{Chandra}. It should be noted here that while clusters are good fields for submillimeter observations, they are challenging space for X-ray observation because of the bright diffuse emission of clusters (e.g., \citealt{2000MNRAS.315L...8F}). We estimate the X-ray luminosities (or mass-accretion rates), star-formation rates, and stellar masses of this sample, and discuss their evolutional stages in comparison with other X-ray detected samples.

The structure of this paper is as follows. In Section~\ref{section:observation}, we describe the details of sample selection and data reductions. Section~\ref{section:analysis} presents the X-ray spectral analysis and the apectral energy distribution (SED) modelling. Section~\ref{section:discussion} and \ref{section:conclusion} are the discussion and conclusion, respectively. Throughout the paper, we assume a flat universe with $H_0=70.4\ \mathrm{km\ s^{-1}\ Mpc^{-1}}$ and $\Omega_M=0.272$ \citep{2011ApJS..192...18K}. The Chabrier initial mass function (IMF, \citealt{2003PASP..115..763C}) is adopted.

\section{Observations and Data Reduction} \label{section:observation}

\begin{figure*}[htbp]
\epsscale{1.18}
\plotone{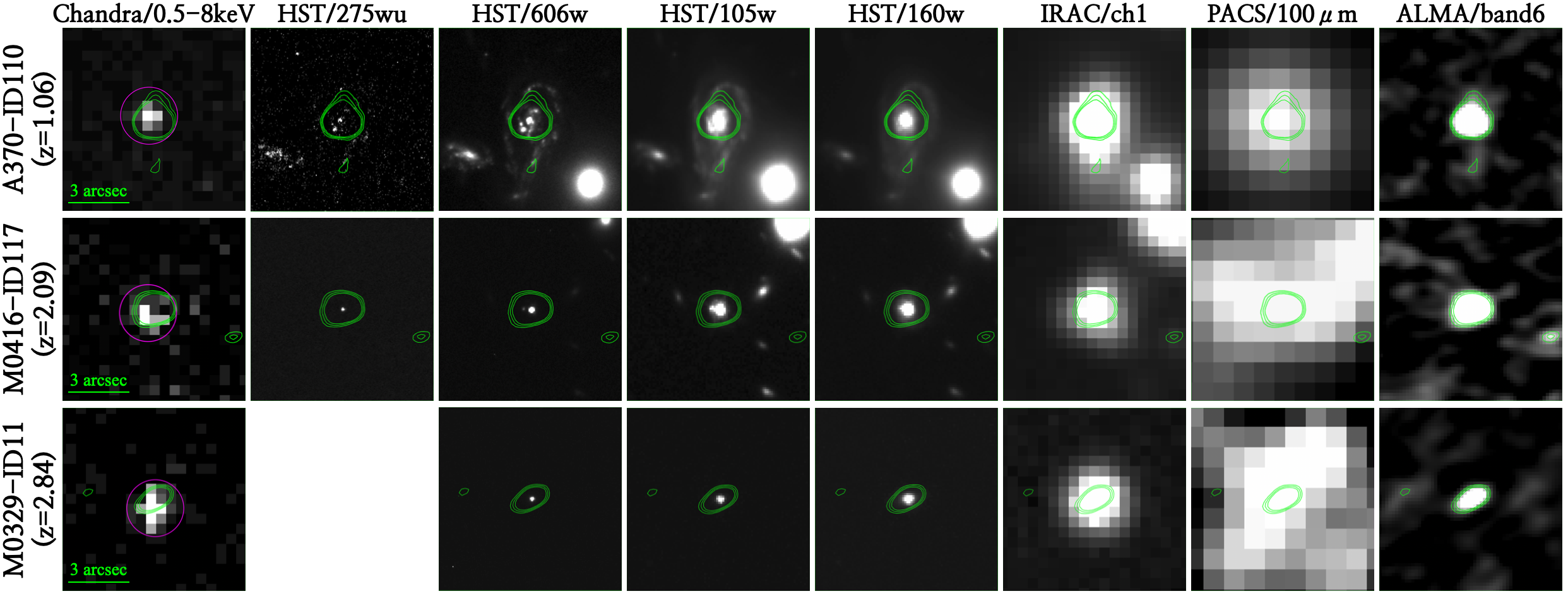}
\caption{
Multi-wavelength images of the ALCS-XAGNs. The green contours show the intensity in the ALMA band drawn at 1$\sigma$ intervals from 3$\sigma$ to 5$\sigma$. The magenta circles in the \textit{Chandra} images show the absolute astrometry uncertainty of \textit{Chandra} (1\farcs4). For the three sources, the statistical error of \textit{Chandra} is much smaller than the absolute one (${<}0\farcs3$). The \textit{Herschel}/PACS images are smoothed with 2D Gaussians of $\sigma$ radii of 1.5 pixels. 
\label{figure:photometry}}
\end{figure*}

\subsection{Observations and Source Detection with ALMA} \label{subsection:alma}

ALMA Band-6 observations of the 33 lensing clusters were conducted between 2018 December and 2019 December through the program 2018.1.00035.L (PI: K.~Kohno; Kohno et al., in prep.) with a 15 GHz total bandwidth covering 250.0--257.5 GHz and 265.0--272.5 GHz. Two compact array configurations (C43-1 and C43-2) were used to obtain a moderate synthesized beam size ($\sim$1\arcsec) to avoid losing sensitivity for spatially extended sources, which can be expected for highly magnified cases. For five Hubble Frontier Fields, we also combined the existing ALMA data from 2013.1.00999S and 2015.1.01425.S (PI: F. Bauer; \citealt{2017A&A...597A..41G,2017A&A...608A.138G}). 

All the ALMA data were calibrated and reduced with the Common Astronomy Software Applications package (CASA; \citealt{2007ASPC..376..127M}) with the pipeline script in the standard manner. Different pipeline versions were used for observations obtained in different cycles, i.e., v5.4.0 for 26 clusters observed in Cycle 6 and v5.6.1 for the remaining clusters in Cycle 7. The natural-weighted 1.15 mm continuum images have a typical noise level of $\sim$60 $\mu$Jy beam$^{-1}$ with a native beam size of $\sim$1\arcsec\ with the natural weighting. We also produced lower resolution maps by applying a $uv$-taper, yielding a typical beam size of $\sim$2\arcsec, which is better suited for spatially elongated, low surface brightness emission. 

The continuum sources were extracted from the native and tapered maps with the natural weighting using the \texttt{SExtractor} version 2.5.0 \citep{1996A&AS..117..393B}. Details of the ALMA data analysis, source extraction processes, and the ALCS 1.2 mm continuum source catalog will be presented in a separate paper (Fujimoto et al. in prep.).

\subsection{Chandra Counterparts} \label{subsection:x-ray}

\textit{Chandra} \citep{2002PASP..114....1W} observed the ALCS fields on multiple occasions since 1999 with the Advanced CCD Imaging Spectrometer (ACIS: \citealt{2003SPIE.4851...28G}). It covers all the ALCS fields except for RXC J0032.1+1808 and RXC J0600.1-2007, with a median exposure of ${\sim}80$ ksec. 155 sources out of the 180 secure ALCS sources are covered by the existing \textit{Chandra} data. We processed all the data obtained by 2017, following the standard analysis procedures with the \textit{Chandra} interactive analysis of observations (CIAO v4.12) software and calibration database (CALDB v4.9.1). The products were combined by using \texttt{merge\_obs} and sources were detected by running \texttt{wavdetect} \citep{2002ApJS..138..185F} in the 0.5--7.0 keV band. In the detection, the false-positive probability threshold was set to $10^{-8}$ and the wavelet scales were set to ``$\sqrt{2}$sequence" (i.e., 1, $\sqrt{2}$, 2, 2$\sqrt{2}$, 4, $\sqrt{2}$, 8, 8$\sqrt{2}$ and 16 pixels), which are the same settings as in \citet{2008ApJS..179...19L}. We cross-matched the \textit{Chandra} detected sources with the secure ALCS source list comprising of 180 millimeter detected galaxies (see Fujimoto et al., in prep.). The X-ray counterpart of an ALMA source was identified if the position coincides with the ALMA position within the \textit{Chandra} positional uncertainty, for which we adopt the root sum square of the 3$\sigma$ statistical error and the 99\% absolute astrometry uncertainty of \textit{Chandra} (1\farcs4).\footnote{\url{https://cxc.harvard.edu/cal/ASPECT/celmon/}} We ignored the positional uncertainty of ALMA, which is much smaller than that of \textit{Chandra}. The cluster center regions are excluded because of the difficulty in identifying point sources.

The three X-ray counterparts were found in the cluster fields of Abell370, MACSJ0416.1--2403, and MACS0329.7--0211, where the total exposure times of the \textit{Chandra} observations are 96.3 ks, 328 ks, and 77.5 ks, respectively. The source names are A370-ID110, M0416-ID117, and M0329-ID11 (hereafter ``ALCS-XAGN''). Their coordinates are summarized in Table~\ref{table:SED}. We extracted the X-ray spectra from circular regions with radii of 1\farcs5 for A370-ID110 and of 2\farcs0 for M0416-ID117 and M0329-ID11\footnote{Since the point spread function (PSF) size at M0416-ID117 and M0329-ID11 is slightly larger than that at A370-ID110, we adopted a little larger aperture for the former two sources.} centered at the X-ray source positions, and subtracted the backgrounds taken from source-free regions around the objects.

It should be noted that these AGNs were also reported in previous studies. A370-ID110 was first discovered by the submillimeter survey of Abell 370 using the Submillimeter Common-User Bolometer Array (SCUBA) \citep{1997ApJ...490L...5S}, and was subsequently detected in X-rays by a \textit{Chandra} follow-up \citep{2000ApJ...543L.119B}. Its properties were discussed in several papers (e.g., \citealt{1999A&A...343L..70S,2002MNRAS.331..495S,2005ApJ...632..736A}). M0416-ID117 was previously detected with ALMA in \citet{2017A&A...597A..41G}, and its physical properties were discussed in \citet{2017A&A...604A.132L}. M0416-ID117 and M0329-ID11 were detected in blind X-ray surveys using \textit{Chandra} archival data by \citet{2016ApJS..224...40W} and \citet{2009MNRAS.392.1509G}, respectively. In this paper, we reanalyze the X-ray spectra and then perform X-ray to millimeter SED analyses utilizing the newly obtained ALMA data.

\subsection{HST, Spitzer, and Herschel Counterparts}

Optical, near-infrared, and mid-infrared images of these three clusters with the ALCS-XAGNs have been taken using \textit{HST} ACS and WFC3, and Spitzer Space Telescope (\textit{Spitzer}) IRAC \citep{2004ApJS..154...10F}. The images and source catalogs were built by reprocessing of available archival \textit{HST} and \textit{Spitzer}/IRAC 3.6 $\mu$m and 4.5 $\mu$m mosaics of ALCS fields \citep{2022arXiv220707125K}. All three ALCS-XAGN sources have counterparts in ACS, WFC3, and IRAC (channel 1 and 2) images as shown in Figure~\ref{figure:photometry}. We also utilized IRAC (channel 3 and 4) and MIPS \citep{2004ApJS..154...25R} photometric data from the Spitzer Enhanced Imaging Products (SEIP; \citealt{https://doi.org/10.26131/irsa433}). Only the counterpart of A370-ID110 was found in the SEIP catalog within 1\farcs5. M0416-ID117 and M0329-ID11 were found outside the coverage of IRAC ch3, ch4 and MIPS.

Herschel Space Observatory (\textit{Herschel}) PACS and SPIRE images at 100--500 $\mu$m bands have been obtained for the ALCS fields (mostly through the \textit{Herschel} Lensing Survey, \citealt{2010A&A...518L..12E}; \citealt{2021ApJ...908..192S}). One of the fields, Abell 370 has been observed as part of the PACS Evolutionary Probe (PEP, \citealt{2011A&A...532A..90L}). Although some of the ALCS fields have only shallow (``snap-shot'') SPIRE coverages without PACS observations, all three ALCS-XAGN containing fields have deep SPIRE and PACS images. The PACS 100 $\mu$m images of these 3 ALCS-XAGN sources are also presented in Figure 1. Details of the \textit{Herschel} data in ALCS fields have been described in \citet{2022ApJ...932...77S}. 

\section{Spectral Analysis and Results} \label{section:analysis}

\subsection{X-ray Spectral Analysis} \label{section:xspectrum}

We fit the observed (magnification not corrected) X-ray spectra with a simple absorbed power-law model. This model is represented as follows in the XSPEC \citep{1996ASPC..101...17A} terminology:
\begin{equation}
\mathrm{\texttt{phabs}*\texttt{zphabs}*\texttt{zpowerlaw}}
\end{equation}
The first term (\texttt{phabs}) represents the galactic photoelectric absorption and the second term (\texttt{zphabs}) an intrinsic absorption at the source redshift. All the parameters are left as free parameters except for the galactic absorption column density and the redshift. The former is fixed at the values estimated by the method of \citet{2013MNRAS.431..394W}. The redshifts are fixed at the spectroscopic ones obtained by Canada-France-Hawaii Telescope in A370-ID110 ($z=1.06$; \citealt{1999A&A...343L..70S}) and the Grism Lens-Amplified Survey from Space in M0416-ID117 ($z=2.09$; \citealt{2016ApJ...831..182H}), and at the photometric one estimated by the optical to near-infrared SED analysis with the \texttt{EAZY} code in M0329-ID11 ($z=2.84^{+0.10}_{-0.11}$; \citealt{2022arXiv220707125K}).

We are able to adequately reproduce the X-ray spectra of the AGNs, with reduced chi-square values of less than 1 ($\chi^2/\mathrm{d.o.f.}<1$). The left panels of Figure~\ref{figure:SED} show the result of the X-ray spectral fitting. The best fit parameters are summarized in Table~\ref{table:SED}. A370-ID110 shows a significant absorption in the X-ray spectrum. This is consistent with the previous X-ray study by \citet{2000ApJ...543L.119B}. We note that A370-ID110 and M0329-ID11 show especially high X-ray luminosities ($\log L_{\mathrm{X}}/\mathrm{erg\ s^{-1}}\gtrsim44.5$) compared with normal AGNs.

\subsection{SED Modelling with \texttt{CIGALE}}\label{subsection:SED}

We perform multi-component SED modelling of the X-ray to millimeter photometry, where we use the magnification-corrected photometries. We employ the latest version of Code Investigating GALaxy Emission (\texttt{CIGALE} v2022.0; \citealt{2005MNRAS.360.1413B,2009A&A...507.1793N,2019A&A...622A.103B,2020MNRAS.491..740Y,2022ApJ...927..192Y}) to conduct X-ray to millimeter SED modeling by self-consistently considering the energy balance between the UV/optical and IR. The SED modules used for the fitting are as follows. We employ a delayed star-formation history (SFH) model, assuming a single starburst with an exponential decay. The simple stellar population (SSP) is modeled with the stellar templates of \citet{2003MNRAS.344.1000B}, where we assume the \citet{2003PASP..115..763C} IMF. We apply a modified Calzetti starburst attenuation law \citep{2000ApJ...533..682C}, where we also allow steeper curve than the original one. The standard nebular emission model (see \citealt{2011MNRAS.415.2920I}) is also added. For the AGN emission, we adopt the \texttt{SKIRTOR} model \citep{2012MNRAS.420.2756S,2016MNRAS.458.2288S}, a clumpy two-phase torus model. The dust emission is modeled by the dust templates of \citet{2017A.A...602A..46J}. We assume an isotropic X-ray radiation from an AGN, where we fix the photon index of 1.9 as a common value, which is well within the errors derived by the X-ray spectral analysis. Here, we use the 2--8 keV absorption corrected flux derived by the X-ray spectral analysis.\footnote{We adopt the average of the upper and lower errors as the flux error.} The redshifts are fixed at the values noted in Section~\ref{section:xspectrum}. The free parameters are summarized in Appendix~\ref{appendix:freeparameter}. The physical properties are estimated by the Bayesian method, where we adopt log-uniform distributions for the prior probability distributions of star-formation rate (SFR)\footnote{In this paper, the term ``SFR'' refers to the average star-formation rate in the last 10 Myr.}, stellar mass ($M_*$), dust luminosity ($L_{\mathrm{dust}}$), and dust mass ($M_{\mathrm{dust}}$)\footnote{In this paper, we use the terms ``dust luminosity'' and ``dust mass'' to describe those of interstellar dust (i.e., not including those in the AGN torus).}, while uniform distributions are assumed for those of minimum radiation field of interstellar dust ($U_{\mathrm{min}}$), infrared excess ($\mathrm{IRX}=\log L_{\mathrm{dust}}/L_{\mathrm{UV,SF}}$), and power-law index of observed UV slope ($\beta$).

\begin{figure*}[p]
\epsscale{1.15}
\plotone{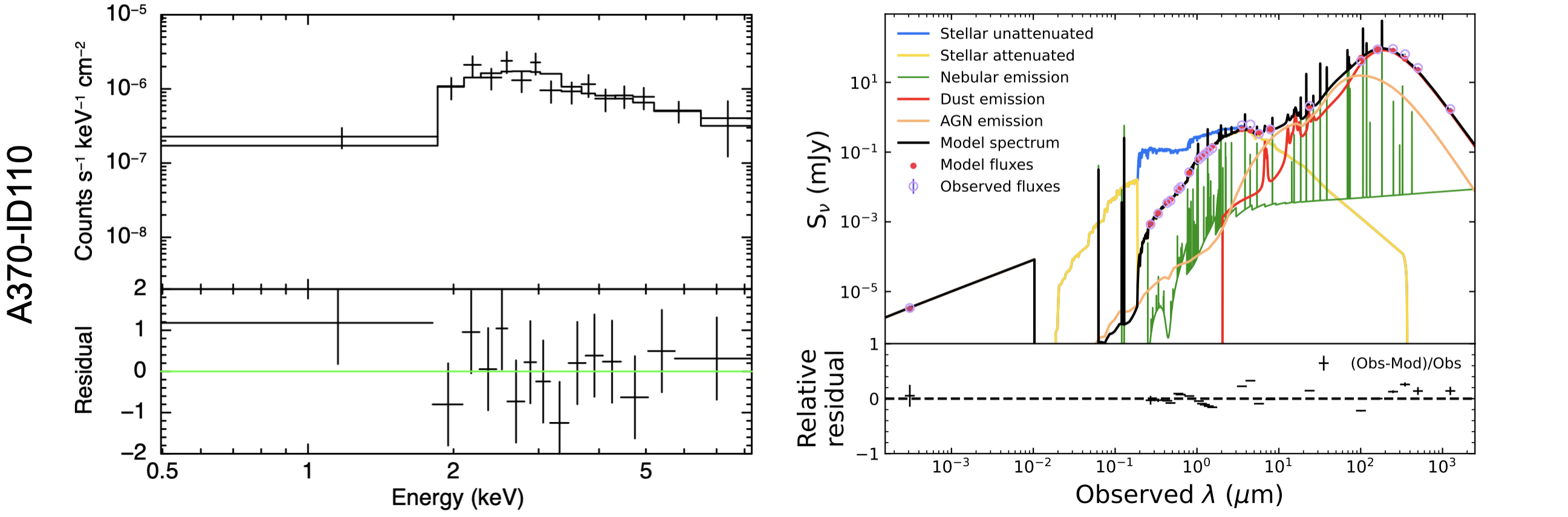}
\plotone{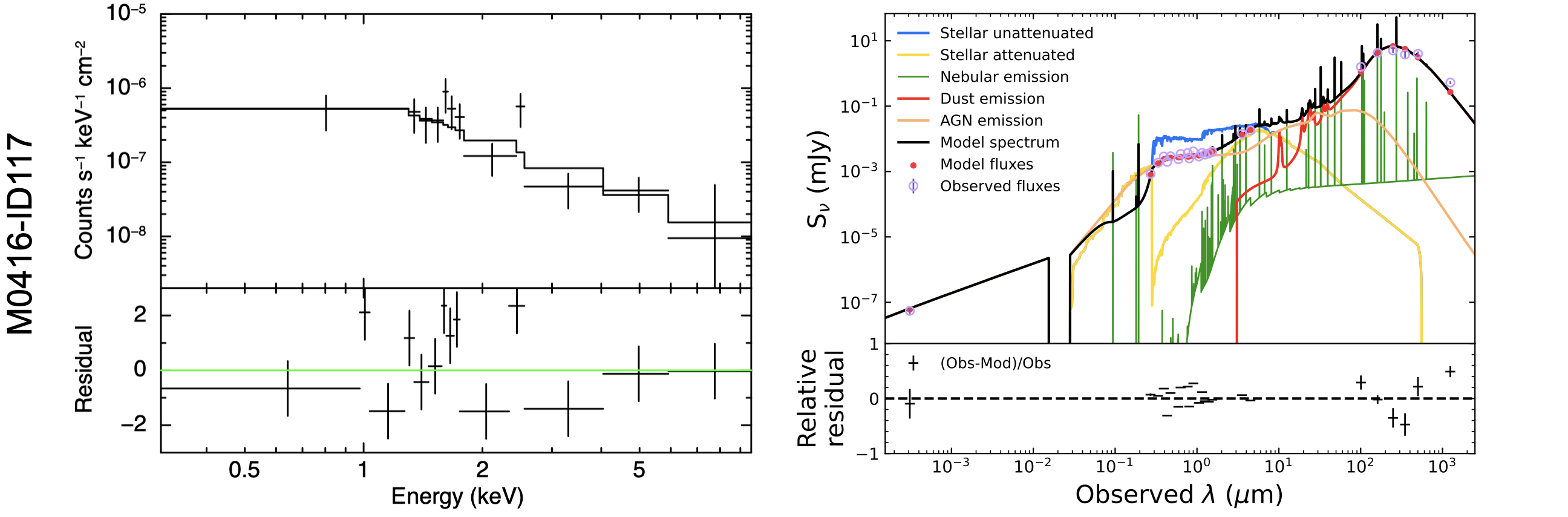}
\plotone{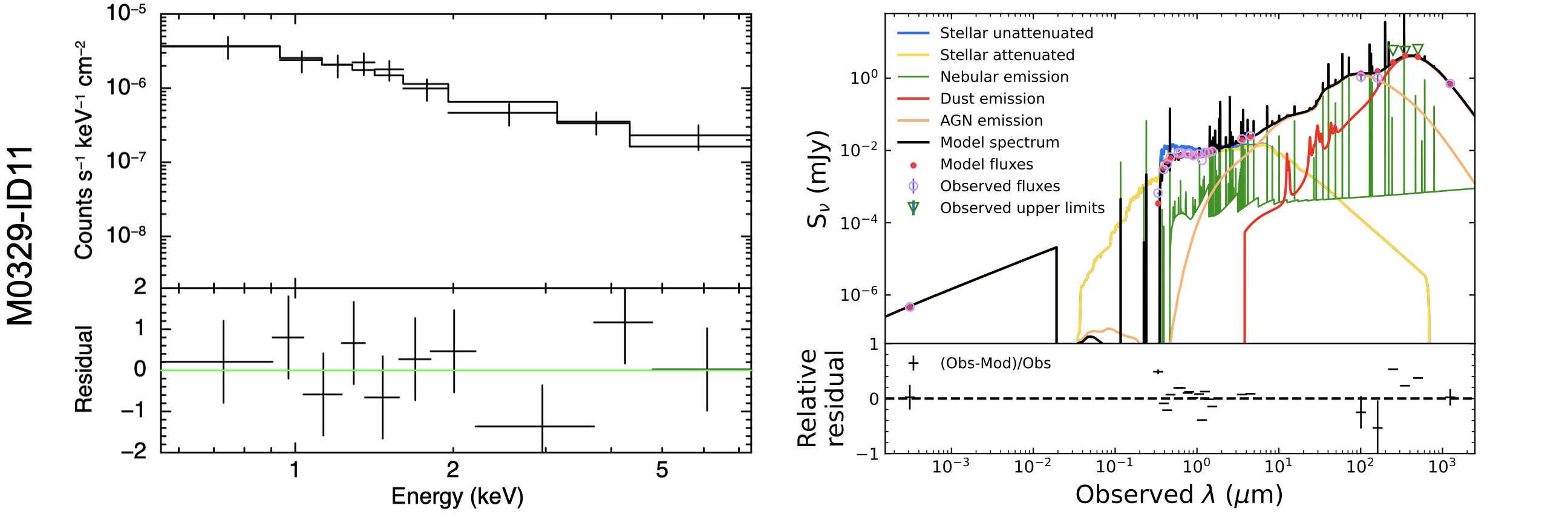}
\caption{
Left: folded X-ray spectra corrected for the effective area and the best fit models of the ALCS-XAGNs. The lower panel shows the residuals. Right: results of the SED modelling. The solid black line represents the best-fit SEDs. The lower panel shows the residuals. The X-ray spectra are not corrected for lensing magnification, while the X-ray to millimeter SEDs are corrected. \label{figure:SED}}
\end{figure*}

We confirm that \texttt{CIGALE} successfully reproduces the SEDs from millimeter to X-ray of all the sources ($\chi^2/\mathrm{d.o.f.}<5$).\footnote{Although this threshold is much larger than in an ordinary chi-square test, we adopt a conservative value by considering the over simplification of the SED model, such as the star-formation history profile, ignorance of time-variability in AGN, and uniform distribution of interstellar dust among stars (see e.g., \citealt{2022A&A...661A..15T}).} The right panels of Figure~\ref{figure:SED} show the results of SED modelling. The physical properties are summarized in Table~\ref{table:SED}. The SED analysis suggests that A370-ID110 is a type 2 AGN, whereas M0416-ID117 and M0329-ID11 are type 1 AGNs. This result is consistent with the AGN types suggested by the X-ray absorption hydrogen column-densities (type 1: $\log N_{\mathrm{H}}/\mathrm{cm^2} \lesssim 22$, type 2: $\log N_{\mathrm{H}}/\mathrm{cm^2} \gtrsim 22$). We also confirm that the near infrared spectrum of A370-ID110 \citep{2006ApJ...651..713T} shows no clear broad emission lines. We find that all the sources are classified as ultra-luminous infrared galaxies ($L_{\mathrm{IR}}/L_{\odot} > 10^{12}$) and show high star-formation rates ($\mathrm{SFR}\gtrsim 100\,M_{\odot}\,\mathrm{yr}^{-1}$). This is consistent with the results presented in \citet{2022ApJ...932...77S}. We also confirm that the physical properties of A370-ID110 and M0416-ID117 are consistent with those reported in the previous studies \citep{2002MNRAS.331..495S,2017A&A...604A.132L}.

\begin{deluxetable*}{lCCCCCCCC}[htb]
\tablecaption{Physical properties of the AGNs and their host galaxies \label{table:SED}}
\tablewidth{\linewidth}
\tablehead{
\colhead{name} & \colhead{R.A.} & \colhead{Decl.} & \colhead{$z$} & \colhead{$\mu$} & \colhead{$N_{\mathrm{H}}$} & \colhead{$\Gamma$} &  \colhead{$\log L_X$} & \colhead{$\log \mathrm{SFR}$}\\[-0.2cm]
& \colhead{(degree)} & \colhead{(degree)} &  &  & \colhead{$(10^{22}\ \mathrm{cm}^{-2})$} &  & \colhead{$(\mathrm{erg\ s^{-1}})$} & \colhead{$(M_{\odot}\ \mathrm{yr^{-1}})$}\\[-0.2cm]
\colhead{(1)} & \colhead{(2)} & \colhead{(3)} & \colhead{(4)} & \colhead{(5)} & \colhead{(6)} & \colhead{(7)} & \colhead{(8)} & \colhead{(9)}\\[-0.1cm]
& \colhead{$\log M_*$} & \colhead{$\log L_{\mathrm{dust}}$} & \colhead{$\log M_{\mathrm{dust}}$} & \colhead{$U_{\mathrm{min}}$} & \colhead{IRX} & \colhead{$\beta$} & \colhead{$f_{\mathrm{AGN}}$} & \colhead{$\log L_{\mathrm{IR}}$}\\[-0.2cm]
& \colhead{$(M_{\odot})$} & \colhead{$(L_{\odot})$} & \colhead{$(M_{\odot})$} &  &  &  &  & \colhead{$(L_{\odot})$}\\[-0.2cm]
& \colhead{(10)} & \colhead{(11)} & \colhead{(12)} & \colhead{(13)} & \colhead{(14)} & \colhead{(15)} & \colhead{(16)} & \colhead{(17)}
}
\startdata
A370-ID110 & 39.985689 & -1.5739784 & 1.06 & 1.19 & 32.8^{+11.7}_{-10.6} & 2.5^{+0.8}_{-0.8} & 45.0^{+0.7}_{-0.5} & 2.43\pm0.02\\
&  11.81\pm0.06 & 12.51\pm0.02 & 8.80\pm0.02 & 25.0\pm0.4 & 2.14\pm0.02 & 0.49\pm0.13 & 0.23\pm0.09 & 12.68\\
M0416-ID117 & 64.0449843 & -24.0798771 & 2.09 & 1.57 & 2.4^{+8.6}_{-2.4} & 2.1^{+0.9}_{-0.6} & 43.8^{+0.6}_{-0.3} & 1.93\pm0.05\\
& 10.81\pm0.16 & 12.01\pm0.03 & 7.98\pm0.11 & 51.2\pm12.5 & 0.84\pm0.03 & -1.85\pm0.02 & 0.05\pm0.01 & 12.07\\
M0329-ID11 & 52.4239873 & -2.1824407 & 2.84 & 2.46 & 0.1^{+5.9}_{-0.1} &  1.7^{+0.6}_{-0.4} & 44.4^{+0.3}_{-0.2} & 2.18\pm0.08\\
& 10.74\pm0.21 & 11.77\pm0.15 & 8.63\pm0.15 & 7.3\pm3.7 & -0.03\pm0.14 & -2.03\pm0.08 & 0.55\pm0.22 & 12.40\\
\enddata
\tablecomments{
(1) Source names. (2)\&(3) ALMA source position. (4) Redshift. Those of A370-ID110 and M0416-ID117 are the spectroscopic redshifts, while that of M0329-ID11 is the photometric redshift (see Section~\ref{section:xspectrum} for more detail). (5) Magnification factor due to the lensing effect (Fujimoto et al., in prep.). (6) X-ray absorption hydrogen column density in units of $10^{22}\ \mathrm{cm}^{-2}$. (7) Photon index. (8) Intrinsic X-ray luminosity in the rest-frame 0.5-8.0 keV band. (9) Star-formation rate. (10) Stellar mass. (11) Total infrared emission from interstellar dust.  (12) Dust mass. (13) Minimum radiation field illuminating the interstellar dust. (14) Infrared excess ($\mathrm{IRX}=\log L_{\mathrm{dust}}/L_{\mathrm{UV}}$). In \texttt{CIGALE}, IRX is calculated from the GALEX FUV filter and dust luminosity. (15) Power-law index of observed UV slope, which is measured in the same way as \citet{1994ApJ...429..582C}. (16) Fraction of AGN IR luminosity to total IR luminosity, where we use the rest-flame 8--1000 $\mu$m luminosities. (17) IR luminosity in the rest-frame 8--1000 $\mu$m band, which is measured by integrating the best-fit SED over a wavelength of 8--1000 $\mu$m.\\
Errors attached in (6)--(8) correspond to the 90\% confidence regions, whereas those in (9)--(16) show the $1\sigma$ confidence regions.\\
All physical quantities in this table are corrected for the lensing magnification; possible uncertainties in the magnification factors are ignored.\\
Because of the poor far-infrared data of M0329-ID11, the IR luminosity may have a large uncertainty.
}
\end{deluxetable*}

\section{Discussions} \label{section:discussion}

We discuss our results in comparison with other ALMA and \textit{Chandra} selected AGN samples (i.e., ALESS-XAGN, UDF-XAGN, ASAGAO-XAGN, and AS2UDS-XAGN) at z=1--3. We also compare our results with X-ray-selected broad line AGNs at $z=$1.18--1.68 in the SXDF (hereafter SXDF-XAGN; \citealt{2008ApJS..179..124U}) as a sample with a different selection method. The stellar masses and SFRs of ALESS-XAGNs, UDF-XAGNs, ASAGAO-XAGNs, and AS2UDS-XAGN are estimated by the \texttt{MAGPHYS} code \citep{2015ApJ...806..110D,2020PASJ...72...69Y,2020MNRAS.494.3828D}, while those of SXDF-XAGNs are estimated by the \texttt{X-CIGALE} code \citep{2021ApJ...909..188S}. The X-ray luminosities of ALESS-XAGNs, UDF-XAGNs and ASAGAO-XAGNs, and AS2UDS-XAGNs are taken from \citet{2013ApJ...778..179W}, \citet{2018ApJ...853...24U}, and \citet{2018ApJS..236...48K}, respectively. We also extract the X-ray luminosities of SXDF-XAGNs from \citet{2012ApJ...761..143N}, where we convert the 2--10 keV luminosity to 0.5--8 keV luminosity, assuming a power-law spectrum with $\Gamma=1.9$. Here we note that all the X-ray luminosities of AS2UDS-XAGNs and ALESS-AGNs are estimated at their spectroscopic redshifts, while the stellar masses and SFRs are estimated at their photometric ones. In this paper, we only plot the sources whose photometric redshifts are close to their spectroscopic ones within $-0.3 < (z_{\mathrm{photo}}-z_{\mathrm{spec}})/(1 + z_{\mathrm{spec}}) < 0.3$. Hence, the systematic errors in stellar masses and SFRs are estimated to be $\sim$0.3 dex, which does not affect our discussions.

Recently, \citet{2019A&A...621A..51H} have investigated the systematic differences among the three SED models: \texttt{GRASIL} \citep{1998ApJ...509..103S}, \texttt{MAGPHYS}, and \texttt{CIGALE}, utilizing the far-ultraviolet to submillimeter SEDs of the 61 galaxies from the KINGFISH sample \citep{2011PASP..123.1347K}. They have shown that galaxies with high specific SFRs ($\log \mathrm{sSFR}[\mathrm{yr}^{-1}]>-10.6$) show good agreement in their stellar masses and SFRs among the three models, whereas those with low sSFRs sometimes show large differences in their SFRs. The sSFRs of the ALESS-XAGNs, UDF-XAGNs, ASAGAO-XAGNs, and AS2UDS-XAGNs are adequately high ($\log \mathrm{sSFR}[\mathrm{yr}^{-1}]>-9.6$), and hence we ignore the systematic difference between \texttt{CIGALE} and \texttt{MAGPHYS} as an approximation.\footnote{The systematic difference between \texttt{CIGALE} and \texttt{MAGPHYS} in the ALCS sample will be discussed in a forthcoming paper on the SED analysis of all the ALCS sources (Uematsu et al., in prep.).} 

\subsection{Stellar Mass versus SFR} \label{subsection:Ms-SFR}

The left panel of Figure~\ref{figure:discussion} shows stellar masses versus SFRs for our 3 sources and other AGN samples (ALESS-XAGN, UDF-XAGN, ASAGAO-XAGN, AS2UDS-XAGN, and SXDF-XAGN). We also show the star-forming ``main sequence'' at $z$=1.0, $z$=2.0, and $z$=3.0 given by \citet{2014ApJS..214...15S}. Our sample is distributed close to the main sequence lines (within 0.2 dex along the SFR axis). This indicates that our sources are normal star-forming galaxies at $z$=1--3, i.e. galaxies with no clear evidence for negative feedback by the AGNs. We note that the ALCS-XAGN sample contains a massive star-forming galaxy (A370-ID110; $M_*>10^{11} M_{\odot}$) and that the SFRs of ALCS-XAGN tend to be smaller than those of the compared SMG samples (see the next section).

\subsection{X-ray luminosity versus SFR} \label{subsection:X_SFR}

\begin{figure*}[htbp]
\epsscale{1.15}
\plotone{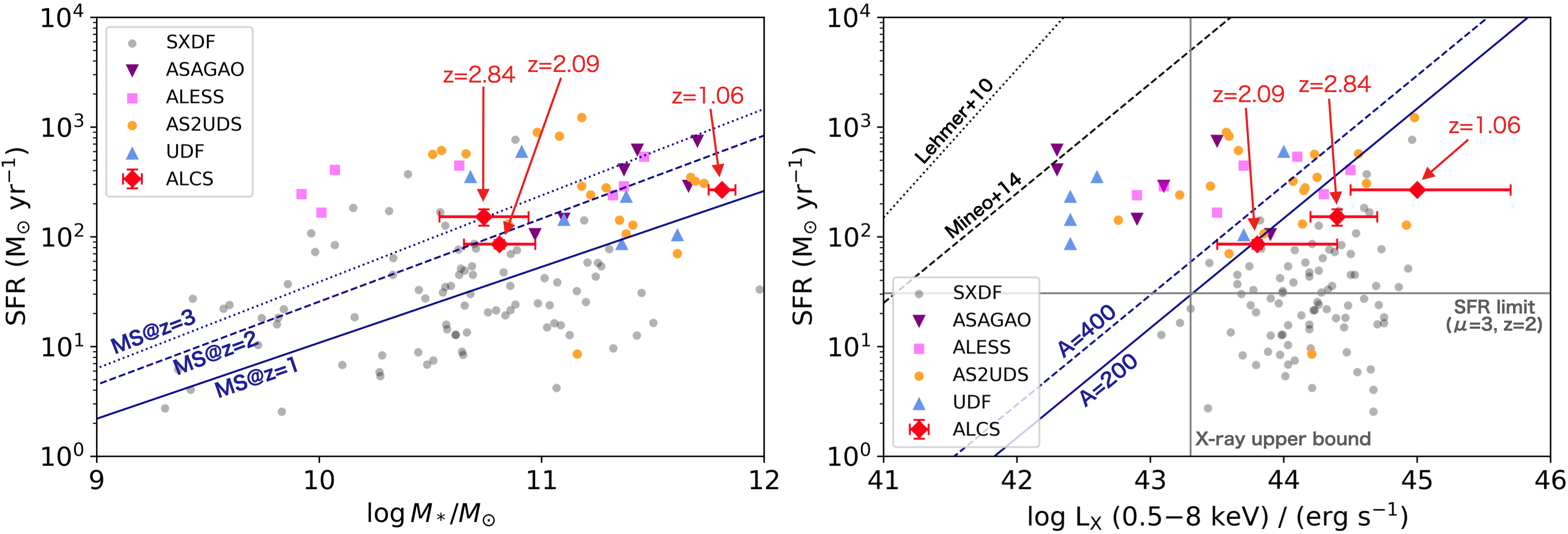}
\caption{
Left: stellar masses ($M_*$) versus SFRs. The blue lines represents the star-forming ``main sequence" at z=1.0 (solid), z=2.0 (dashed), and z=3.0 (dotted), respectively \citep{2014ApJS..214...15S}. Right: de-absorbed X-ray luminosities versus SFRs. The solid and dashed blue lines represent the relation of galaxy-SMBH simultaneous evolution for A=200 and A=400, respectively (Section~\ref{subsection:X_SFR}). All the physical quantities in this figure are corrected for lensing magnification. For comparison, we show the X-ray contribution from star-formation activity with dotted \citep{2010ApJ...724..559L} and dashed \citep{2014MNRAS.437.1698M} black lines. We also show the median X-ray upper bounds of \textit{Chandra} non-detected ALCS sources with vertical grey line, where we assume an intrinsic X-ray absorption of $\log N_{\mathrm{H}}/\mathrm{cm}^{-2}=23$ (Uematsu et al., in prep.). Moreover, we show the detection limit of SFR with horizontal grey line. The SFR limit is converted from the infrared luminosity limit by assuming the Kennicutt law \citep{1998ARA&A..36..189K}, where we multiply 0.63 to correct Salpeter IMF to Chabrier IMF. The infrared luminosity limit is estimated from the observed flux limit at 1.2 mm (${\sim}$300 $\mu$Jy; 5$\sigma$) by assuming a single greybody with an emissivity index of 1.8 and temperature of 35 K at $z=2$, using a typical magnification factor of 3.
\label{figure:discussion}}
\end{figure*}

The right panel of Figure~\ref{figure:discussion} shows X-ray luminosities versus SFRs. We also show the relation of galaxy-SMBH simultaneous evolution for the local $M_{\mathrm{BH}}$-vs-$M_{\mathrm{bulge}}$ and $M_{\mathrm{BH}}$-vs-$M_{\mathrm{stellar}}$ relations. This relation is given as
\begin{equation}
\mathrm{SFR}*(1-R)=A\times \dot{M}_{\mathrm{BH}}
\end{equation}
where $R$ is the return fraction ($R=0.41$ for a Chabrier IMF), and $A$ is the mass ratio ($M_{\mathrm{bulge}}/M_{\mathrm{BH}}=200$ and $M_{\mathrm{stellar}}/M_{\mathrm{BH}}=400$; see \citealt{2018ApJ...853...24U} for more details). The mass-accretion rate can be estimated by the X-ray luminosity as
\begin{equation}
\dot{M}_{\mathrm{BH}}=\kappa_{0.5-8}L_{\mathrm{X}}(1-\eta)/(\eta c^2)
\end{equation}
where $\kappa_{0.5-8}$ is the bolometric correction factor ($\kappa_{0.5-8}=13$ or $\kappa_{2-10}=20$ assuming a photon index of 1.9; \citealt{2007MNRAS.381.1235V}), $\eta$ is the radiation efficiency ($\eta=0.05$), and $c$ is the speed of light. ALCS-XAGNs are located on the ``simultaneous evolution'' line or in the AGN dominant phase, while the other ALMA and \textit{Chandra} selected samples belong to the SF-dominant phase. This is a unique feature of our ALCS-XAGN sample among sub/millimeter selected galaxies, realized by the following selection effects. Since the X-ray data of ALCS are shallower but cover a wider area than those of the UDF survey and ASAGAO, high X-ray luminosity (hence rare) AGNs can be efficiently selected in the ALCS sample. On the other hand, because ALCS is much deeper than the parent sample of ALESS and AS2UDS\footnote{Since ALESS and AS2UDS are follow-up observations of the single-dish surveys LESS \citep{2009ApJ...707.1201W} and S2CLS \citep{2017MNRAS.465.1789G}, respectively, their properties are basically determined by the selection bias of the parent samples.}, whose median SFRs are ${\sim}250\,M_{\odot}\,\mathrm{yr}^{-1}$, ALCS can detect AGNs in the AGN-dominant phase, which are relatively faint in the millimeter band with SFRs $\lesssim 250\,M_{\odot}\,\mathrm{yr}^{-1}$. Besides, since the ALCS-XAGNs are selected by millimeter observations, they are more biased to higher SFRs compared with purely X-ray selected AGN samples (e.g., SXDF-XAGNs whose median SFR is ${\sim}30\,M_{\odot}\,\mathrm{yr}^{-1}$). \citet{2012Natur.485..213P} showed that submillimeter galaxies containing AGNs with high mass-accretion rates show relatively low SFRs, implying negative feedback by the AGNs. Given this trend, A370-ID110 and M0329-ID11, which show both high SFRs ($\mathrm{SFR}>100\,M_{\odot}\,\mathrm{yr}^{-1}$) and high X-ray luminosities ($\log L_{\mathrm{X}}/\mathrm{erg\ s^{-1}}\gtrsim44.5$), may be a rare population that is difficult to be detected in previous sub/millimeter surveys.

According to the merger-driven evolutionary scenario proposed by many authors (e.g., \citealt{1996ARA&A..34..749S,2008ApJS..175..356H,2021ApJS..257...61Y,2022ApJ...936..118Y}), the galaxies and SMBHs ``co-evolve'' during major mergers being deeply embedded by gas and dust, and then evolved to the AGN dominant phase with quenched SFRs by AGN feedback. If this scenario is applicable, our sources may correspond to the transition stage where the merging has finished, but the star formation is not yet quenched. The smaller obscuration in our AGNs than that in late-stage mergers \citep{2017MNRAS.468.1273R,2021MNRAS.506.5935R,2021ApJS..257...61Y} is also consistent with this picture.

We note, however, that it is not clear whether the AGNs and star-formation activities in our sample are indeed triggered by merger processes or not. Some studies argue that a significant fraction of AGNs at $z\sim$1--3 are triggered by secular mechanisms (e.g., \citealt{,2011ApJ...727L..31S,2012ApJ...744..148K,2012ApJ...751...72D}), where star-formation activities may inject turbulence in the gas disks and make the gas fall into the nuclear regions \citep{2011MNRAS.413.2633H}. Given the high SFRs of ALCS-XAGNs, the AGN activities might be triggered by the intense star-formation activities in their host galaxies.

\citet{2018ApJ...861....7F} studied the morphology of dusty star-forming galaxies detected in ASAGAO at $z$=1--3. They found that the ULIRGs at $z$=1--3 showed larger fractions of irregular and merging galaxies than less luminous galaxies, implying the significant contribution of ongoing merger process in high-redshift ULIRGs, although the fraction of disk galaxies is the largest even in the ULIRG sample. In our X-ray AGN sample, only A370-ID110 is confirmed to have disk-like structure, while M0416-ID117 and M0329-ID11 are too faint to examine their morphology. Future investigation of host galaxy morphology with James Webb Space Telescope or Thirty Meter Telescope would be helpful to reveal the origins of the high mass accretion rates and SFRs in our sample.

\section{conclusions} \label{section:conclusion}

We have reported the multi-wavelength properties of millimeter galaxies hosting AGNs detected in the ALCS. The main conclusions are summarized as follows:

\begin{enumerate}

\item Utilizing the archival data of \textit{Chandra}, we have identified 3 AGNs out of the 180 millimeter galaxies securely detected in the ALCS, which are found in the fields of Abell370, MACSJ0416.1--2403, and MACS0329.7--021 at spectroscopic redshifts of 1.06 and 2.09, and photometric redshift of 2.84, respectively.

\item The X-ray spectral analysis shows that two AGNs are not significantly absorbed ($\log N_{\mathrm{H}}/\mathrm{cm}^{-2} < 23$), while one shows signs of moderate absorption ($\log N_{\mathrm{H}}/\mathrm{cm}^{-2} \sim 23.5$).

\item We have performed SED modelling of the X-ray to millimeter photometry with the \texttt{CIGALE} code. We find that our sources have both high mass-accretion rates (intrinsic 0.5--8 keV X-ray luminosities of ${\sim}10^{44-45}\ \mathrm{erg\ s^{-1}}$) and star-formation rates (${\gtrsim}100M_{\odot}\,\mathrm{yr}^{-1}$).

\item We find that ALCS-XAGNs show higher mass-accretion rates than other ALMA selected X-ray AGN samples. We also confirm that ALCS-XAGNs show higher SFRs than a purely X-ray-selected AGN sample. This can be explained by the selection bias, showing that a wide and deep survey with ALMA, combined with medium-depth X-ray data, can efficiently detect intense growth stage of both galaxies and SMBHs in high-redshift universe.

\end{enumerate}

\clearpage
\acknowledgments
We thank the anonymous referee for providing useful comments to improve the quality of the paper. We thank Ian Smail for very helpful discussions. This publication uses data from the ALMA programs: ADS/JAO.ALMA\#2018.1.00035.L, \#2013.1.00999.S, and \#2015.1.01425.S. ALMA is a partnership of ESO (representing its member states), NSF (USA) and NINS (Japan), together with NRC (Canada), MOST and ASIAA (Taiwan), and KASI (Republic of Korea), in cooperation with the Republic of Chile. The Joint ALMA Observatory is operated by ESO, AUI/NRAO and NAOJ. This work has been financially supported by JSPS KAKENHI Grant Numbers 22J22795 (R.U.), 20H01946 (Y.U.), 17H06130 (K.K., Y.U.), 19K14759 (Y.T.), and 22H01266 (Y.T.), and the NAOJ ALMA Scientific Research Grant Number 2017-06B (K.K.). S.Y. is grateful for support from RIKEN Special Postdoctoral Researcher Program. F.S. acknowledges support from the NRAO Student Observing Support (SOS) award SOSPA7-022. G.E.M. acknowledges the Villum Fonden research grants 13160 and 37440 and the Cosmic Dawn Center of Excellence funded by the Danish National Research Foundation under the grant No. 140. HU acknowledges support from JSPS KAKENHI Grant Number 20H01953.

\vspace{5mm}
\facilities{ALMA, \textit{Chandra}, \textit{HST}, \textit{Spitzer}, \textit{Herschel}}

\software{HEAsoft v6.27 \citep{2014ascl.soft08004N},  
   CIAO v4.12 \citep{2006SPIE.6270E..1VF}, 
   CIGALE v2022.0 \citep{2022ApJ...927..192Y},
   CASA v5.4.0 \citep{2007ASPC..376..127M},
   SExtractor v2.5.0 \citep{1996A&AS..117..393B},
   EAZY \citep{2008ApJ...686.1503B}
   }

\clearpage
\appendix

\section{Parameters used in the SED analysis}\label{appendix:freeparameter}

Table~\ref{table:xcigale} summarize the free parameters used in the SED modelling. Each SED module is explained in Section~\ref{subsection:SED}.

\begin{table*}[h]
\caption{Parameter Ranges Used in the SED Modelling with \texttt{CIGALE}.\label{table:xcigale}}
\begin{tabularx}{\linewidth}{@{}lll}
 \hline\hline
 Parameter &\hspace{5pt}Symbol &\hspace{5pt}Value\\
 \hline
 \multicolumn3c{SFH (Delayed SFH)}\\
 \hline
 e-folding time of the main stellar population &\hspace{5pt}$\tau_{\mathrm{main}}$ [Myr] &\hspace{5pt}300, 600, 1000, 3000, 5000, 7000, 9000\\
 Age of the main stellar population &\hspace{5pt}$\mathrm{age_{main}}$ [Myr] &\hspace{5pt}300, 600, 1000, 2000, 3000, 4000, 5000\\
 \hline
 \multicolumn3c{SSP (bc03; \citealt{2003MNRAS.344.1000B})}\\
 \hline
 IMF of the stellar model &\hspace{5pt} &\hspace{5pt}\citet{2003PASP..115..763C}\\
 Metalicity of the stellar model &\hspace{5pt} &\hspace{5pt}0.02\\
 \hline
 \multicolumn3c{Dust Attenuation (dustatt\_modified\_starburst; \citealt{2000ApJ...533..682C})}\\
 \hline
 The colour excess of the nebular lines. &\hspace{5pt}$E(B-V)_{\mathrm{lines}}$ &\hspace{5pt}0.05, 0.1, 0.15, 0.2, 0.4, 0.6, 0.8, 1.0,\\
 & &\hspace{5pt}1.2, 1.4, 1.6, 1.8, 2.0, 2.2, 2.4\\
 Reduction factor to apply $E(B-V)_{\mathrm{lines}}$ to calculate the stellar &\hspace{5pt}$E(B-V)_{\mathrm{factor}}$ &\hspace{5pt}0.44\\
 continuum attenuation & & \\
 Power-law index to modify the attenuation curve &\hspace{5pt}$\delta$ &\hspace{5pt}-0.6, -0.3, 0.0\\
 \hline
 \multicolumn3c{AGN emission (skirtor2016; \citealt{2012MNRAS.420.2756S,2016MNRAS.458.2288S})}\\
 \hline
 Average edge-on optical depth at 9.7 micron &\hspace{5pt}$\tau_{\rm 9.7}$ &\hspace{5pt}3, 7, 11\\
 Radial gradient of dust density &\hspace{5pt}$p$ &\hspace{5pt}1.0\\
 Dust density gradient with polar angle &\hspace{5pt}$q$ &\hspace{5pt}1.0\\
 Half-opening angle of the dust-free cone &\hspace{5pt}$\Delta$\,[\degr] &\hspace{5pt}40\\
 Ratio of outer to inner radius &\hspace{5pt}$R$ &\hspace{5pt}20\\
 Inclination &\hspace{5pt}$\theta$\,[\degr] &\hspace{5pt}30, 60\\
 Fraction of AGN IR luminosity to total IR luminosity &\hspace{5pt}$f_{\rm AGN}$ &\hspace{5pt}0.1, 0.3, 0.5, 0.7, 0.9\\
 Extinction in polar direction &\hspace{5pt}$E(B-V)_{\mathrm{pol}}$ &\hspace{5pt}0.0, 0.1, 0.8\\
 Temperature of the polar dust &\hspace{5pt}$T_{\mathrm{pol}}$\,[K] &\hspace{5pt}100, 200\\
 \hline
 \multicolumn3c{Dust emission (Themis; \citealt{2017A.A...602A..46J})}\\
 \hline
 Mass fraction of small hydrocarbon solids &\hspace{5pt}$q_{\mathrm{PAH}}$ &\hspace{5pt}2.5\\
 Minimum radiation field &\hspace{5pt}$U_{\mathrm{min}}$ &\hspace{5pt}1, 5, 10, 25, 50\\
 Power-law index of the starlight intensity distribution &\hspace{5pt}$\alpha$ &\hspace{5pt}2.0, 2.5\\
 Fraction of dust illuminated from Umin to Umax &\hspace{5pt}$\gamma$ &\hspace{5pt}0.01, 0.05, 0.1\\
 \hline
 \multicolumn3c{X-ray emission}\\
 \hline
 Photon index &\hspace{5pt}$\Gamma$ &\hspace{5pt}1.9\\
 Maximum deviation of $\alpha_{\mathrm{ox}}$ from the empirical relation\tablenotemark{a} &\hspace{5pt} &\hspace{5pt}0.2\\
 \hline
\end{tabularx}
\tablenotetext{a}{We adopt the well-studied ``$\alpha_{\mathrm{ox}}$-$L_{2500 \mathrm{\mathring{A}}}$" relation, where $\alpha_{\mathrm{ox}}$ is defined as $\alpha_{\mathrm{ox}}\equiv-0.3838\times \log\left(L_{\mathrm{2500 \mathrm{\mathring{A}}}}/L_{\mathrm{2keV}}\right)$ and the empirical relation is written as $\alpha_{\mathrm{ox}}=-0.137\log\left(L_{2500 \mathrm{\mathring{A}}}\right)+2.638$ \citep{2007ApJ...665.1004J}.}
\end{table*}

\bibliography{citation}{}
\bibliographystyle{aasjournal}

\end{document}